\documentclass[twocolumn,floats,prl,aps]{revtex4}

\usepackage{graphicx}
\usepackage{dcolumn} 
\usepackage{amsmath}
\usepackage{exscale}
\usepackage{bm}      
\usepackage{color}
\usepackage{epsfig,psfrag,subfigure,subeqnarray}
\usepackage{bbm}
\usepackage{amsbsy}
\usepackage{amssymb}
\usepackage{enumitem}
\usepackage{mathrsfs}
\usepackage{amsfonts}

\def\lan{\langle}
\def\ran{\rangle}
\def\va{\varepsilon}

\def\vk{{\bf k}}
\def\vK{{\bf K}}

\def\vq{{\bf q}}
\def\vp{{\bf p}}

\def\v0{{\bf 0}}
\newcommand{\bd}{\begin{equation}}
\newcommand{\ed}{\end{equation}}
\newcommand{\be}{\begin{equation}}
\newcommand{\ee}{\end{equation}}
\newcommand{\bt}{\begin{split}}
\newcommand{\et}{\end{split}}

\newcommand{\bn}{\begin{align}}
\newcommand{\en}{\end{align}}

\newcommand{\bea}{\begin{eqnarray}}
\newcommand{\eea}{\end{eqnarray}}
\newcommand{\ba}{\begin{array}}
\newcommand{\ea}{\end{array}}
\newcommand{\nn}{\nonumber}

\DeclareMathAlphabet\mathbfcal{OMS}{cmsy}{b}{n}

\begin{document}

\title{Symmetry breaking for semiconductor excitons 

induced by Coulomb coupling between heavy and light holes}

\author{Shiue-Yuan Shiau$^1$, Benoit Eble$^2$ and Monique Combescot$^2$}\
\affiliation{(1) Physics Division, National Center for Theoretical Sciences, Taipei, 10617, Taiwan}
\affiliation{(2) Institut des NanoSciences de Paris, Sorbonne Universit\'e, CNRS, 75005 Paris, France}


\begin{abstract}
Semiconductor excitons are commonly seen as hydrogen atom. This analogy requires a unique hole mass. In reality, this is not so due to the complexity of the semiconductor band structure. The precise consequences on the exciton physics of the Coulomb coupling between heavy and light holes remain a tricky open problem. Through an ``optimized perturbative'' approach that uses excitons with a flexible hole mass as a basis, we show that for zero exciton wave vector, the heavy-light hole mass difference does not split the $(2\times4)$ exciton degeneracy in zinc-blende-like semiconductors, the hole mass for binding energy being close to the average mass inverse. By contrast, for nonzero exciton wave vector, that physically breaks the crystal symmetry, the exciton degeneracy splits into two branches quantized along the exciton wave vector, with nontrivial center-of-mass dependence not only on the heavy and light hole masses, but also on the electron mass.  
\end{abstract}
\date{\today}
\maketitle


 The impact of excitons on linear and nonlinear optics of semiconductors is of paramount importance\cite{Kirabook} for modern technology. Powered by mature conventional heterostructures and novel low-dimensional materials, a new momentum has recently gathered to exploit excitons for sensing\cite{Feierabend2017}, quantum simulation\cite{Lagoin2022}, quantum memory and quantum circuits\cite{Krenner2008,Cong2020}. 
 
 The exciton is a composite quantum particle usually seen as one negatively-charged conduction electron and one positively-charged valence hole correlated by Coulomb attraction, in this way sharing similarity with the hydrogen atom. Driven by this attraction, bound-state excitons show up as pronounced narrow absorption lines below the band gap. 
 
 Such a simplified exciton picture however has to be questioned when facing cutting-edge challenges in the emerging field of ``Excitonics''\cite{Kis_2019} because the hole definitely is a very tricky quantum object: it fundamentally corresponds to an electron absence in the full valence band, with spin and spatial degeneracies mixed by the spin-orbit interaction. The resulting valence band complexity defies the gross exciton reduction to hydrogen-like eigenstates. Even within the spherical approximation valid for zinc-blende-like cubic semiconductors, there are still two holes with different masses, coupled by Coulomb scatterings in a far from trivial way\cite{Shiauprb21}. These two masses prevent using the hydrogen procedure --- that fundamentally transforms the two-body problem into two separate one-body problems --- to solve the resulting Schr\"odinger equation \textit{exactly}, as required to derive bound states. A procedure, completely different from the very first line, must be found.

The fact that heavy holes can turn light under Coulomb interaction, leads us to anticipate that for \textit{zero} exciton center-of-mass wave vector, the heavy and light hole masses should appear in the exciton binding energy, through a unique averaged value because in the absence of cubic-symmetry breaking, there is no physical reason for the exciton energy to split. By contrast, a \textit{nonzero} exciton wave vector that produces such a symmetry breaking, should lead to a splitting of the exciton energy into two branches with different center-of-mass masses.

\textbf{We here show} that the exciton binding energy is driven by a single hole mass value, so that the exciton degeneracy is preserved despite  the hole mass difference. This single mass is close to $m^\ast_h$, obtained by averaging the hole mass inverses (see Eq.~(\ref{HL_2})), not only for small hole mass difference, but also for light conduction electron. Yet, this single hole mass does not describe the exciton motion: when the exciton wave vector $\vK$ differs from zero, the exciton splits into two branches  in which the heavy and light holes appear through the two ($\pm3/2$) and ($\pm1/2$) linear combinations quantized along $\vK$ in a way similar to the heavy-light hole splitting induced by the spin-orbit interaction in the valence band.

The  major difficulty of this problem is to handle the Coulomb interaction exactly --- as required for bound states --- while the extremely complicated heavy-light Coulomb scatterings\cite{Shiauprb21} deprive any hope to analytically solve the corresponding Schr\"odinger equation. The trick we have found to overcome this difficulty, is to introduce an exciton basis constructed on holes having a single, flexible hole mass $m_h$. Its states are obtained from a hydrogen-like procedure, with the Coulomb interaction handled exactly. We look for the consequences of having two hole masses different from $m_h$, by treating the residual one-body hole kinetic term through a perturbative approach. The $m_h$ hole mass is ultimately adjusted for the resulting exciton ground-state energy to be minimum. The procedure we propose, that can be qualified as an ``optimized perturbative approach'', is similar to the optimized $\delta$ expansion\cite{expansion1,expansion2} that aims at obtaining better than conventional perturbative results. This procedure allows us to catch the physics of the heavy-light hole problem.  The result we find reads in terms of the hole mass difference $(m_{_H}-m_{_L})/(m_{_H}+m_{_L})$, within a $m_e/(m_e+m^\ast_h)$ prefactor, for the electron mass has to show up  in some way when dealing with exciton.

\textit{\textbf{The problem}} --
We  consider a zinc-blende-like cubic semiconductor with valence band  having a threefold spatial degeneracy labeled as $\mu=(x,y,z)$ along the crystal axes. The spin-orbit interaction mixes the $(3\times2)$ hole states $|\mu\ran\otimes|s\ran$ and splits their subspace into a fourfold level $\mathcal{J}_\textbf{z}=(\pm3/2,\pm1/2)$ quantized along the crystal axis $\textbf{z}$, and a twofold level that we here neglect as it is far below in energy\cite{Cardona}. The $\vk\cdot\vp$ coupling of the fourfold valence electrons to the conduction levels\cite{Fishman} produces two energy dispersions associated with heavy and light masses $(m_{_H},m_{_L})$ in the spherical approximation that neglects the warping\cite{DresselhausPR1955,Dresselhaus1956,Luttinger,Baldereschi_prl,Baldereschi_prb}. 
 The resulting heavy and light holes are labeled 
by $\mathcal{J}_{\vk_h}=(\pm3/2)$ and $\mathcal{J}_{\vk_h}=(\pm1/2)$ indices that are quantized along the hole wave vector 
$\vk_h$.

\begin{figure}
\includegraphics[trim=6.6cm 10.5cm 2.9cm 4cm,clip,width=4in]{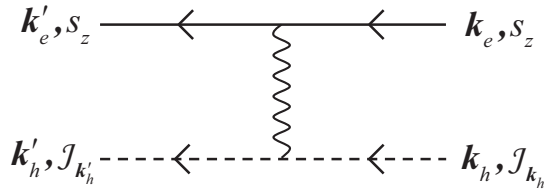}
\caption{A ``in'' heavy hole, $\mathcal{J}_{\vk_h}=\pm3/2$, can stay heavy, $\mathcal{J}_{\vk'_h}=\pm3/2$, or turn light, $\mathcal{J}_{\vk'_h}=\pm1/2$, under a Coulomb interaction. The associated scattering depends on the angle between the ``in''  and ``out'' hole wave vectors $(\vk_h,\vk'_h)$ in a very complicated way\cite{Shiauprb21}. }
\label{fig1}
\end{figure}

The Coulomb interaction, diagonal for holes in the spatial Bloch states labeled by $\mu$, stays diagonal\cite{Monicbook} for spin-orbit holes labeled by $\mathcal{J}_\textbf{z}$, but does not stay diagonal\cite{Shiauprb21} for $\mathcal{J}_{\vk_h}$ holes because the hole wave vector $\vk_h$ changes in a Coulomb scattering (Fig.~\ref{fig1}). This is why working with different heavy and light masses makes the exciton problem in bulk samples tremendously difficult. Note that this difficulty does not exist in quantum wells because the confinement energy brings the light holes far below in energy, so that we can neglect their role\cite{dense ou puits}.

It is then not a surprise that when dealing with bulk exciton, a unique mass is assigned to the hole, its common value\cite{DresselhausPR1955,Dresselhaus1956} reading in terms of heavy and light  masses as 
\be\label{HL_2}
\frac{1}{m^\ast_h}=\frac{1}{2}\left(\frac{1}{m_{_H}}+\frac{1}{m_{_L}}\right)
\ee
While averaging mass inverses is reasonable for exciton binding  because this binding depends on the hole mass as $1/m_h$, such an averaging is hard to accept for exciton center of mass because its hole part would then tend to $2m_{_L}$ for $m_{_H}\gg m_{_L}$.

As the Coulomb interaction couples heavy and light holes, a heavy hole can turn light in a Coulomb scattering, with an amplitude that heavily depends\cite{Shiauprb21} on the angle between the incoming and outgoing wave vectors, $\vk_h$ and $\vk'_h$ (Fig.~\ref{fig1}). For holes having different masses, the exciton that results from the repeated Coulomb scatterings between electron and hole is not analytically solvable for a very simple reason: it is no more possible to reduce the two-body electron-hole problem to two one-body problems. Indeed, the kinetic energy of an electron-hole pair with masses $(m_e,m_h)$ can be rewritten as
\be\label{HL_3}
\frac{\hbar^2\vk_e^2}{2m_e}+\frac{\hbar^2\vk_h^2}{2m_h}=\frac{\hbar^2\vK^2}{2(m_e+m_h)}+\frac{\hbar^2\vk^2}{2}\left(\frac{1}{m_e}+\frac{1}{m_h}\right)
\ee 
for $\vk_e=\vk+\gamma_e\vK$ and $\vk_h=-\vk+\gamma_h\vK$ with $\gamma_e=1-\gamma_h=m_e/(m_e+m_h)$. Depending on which hole is scattered, $m_h$ can be $m_{_H}$ or $m_{_L}$. So, although the  pair center-of-mass wave vector, $\vk_e+\vk_h=\vK$, stays constant in a Coulomb scattering, it is impossible to define a pair relative-motion wave vector $\vk$ and follow its change at each  Coulomb interaction.

To approach this problem, we introduce the $\hat{H}_0$  Hamiltonian in which all $\mathcal{J}_\vk=(\pm3/2,\pm1/2)$ holes are taken with the same mass $m_h$, not necessarily equal to $m^*_h$. The resulting ground-state exciton, obtained from the standard hydrogen-like procedure, is $(2\times4)$-fold degenerate. The consequences of the hole mass difference follow from $\hat{\Delta}=\hat{H}-\hat{H}_0$ acting in this eightfold  exciton subspace\cite{SM}, for $\hat{H}$ being the Hamiltonian when $m_{_H}\neq m_{_L}$. 

\begin{table}[t]
\caption{Values for $m_h^\ast$ in Eq.~(\ref{HL_2}), $m_h^{\ast\ast}$ in Eq.~(\ref{mh**}),  $E^{(\pm)}_\vK=E_\vK(1\pm r)$ in Eq.~(\ref{HL_5}) and $T_K=E_\vK/ k_{_B}$, for various semiconductors. The mass unit is the free electron mass $m_0$. The exciton wave vector $\vK$ is obtained from $E_{gap}=\hbar (c/n_r) |\vK|$ where $n_r$ is the semiconductor refractive index. }
\centering
\label{table1} 
\begin{tabular}{m{2cm} m{1.5cm} m{1.5cm} m{1.5cm} m{1cm}}
\hline\hline
     &  GaAs & GaSb & InP & InAs \\[0.5ex]
  \hline
$m_e$  & 0.063 &0.041 & 0.08 & 0.023\\
 
 $m_{_H}$  & 0.51 &0.4 & 0.6 & 0.42\\
$m_{_L}$  & 0.082 &0.05 & 0.089 & 0.026\\
$E_{gap}$ (eV) & 1.52 & 0.81 & 1.42 & 0.42 \\
$m_h^\ast$ & 0.14  & 0.089 & 0.16 &  0.049 \\

$m_h^{\ast\ast}/m_h^\ast$ & 1.15  &1.17 & 1.17 & 1.22 \\
$r$ & 0.5  & 0.53 & 0.49 & 0.6 \\
$n_r$ & 3.26  & 3.71 & 3.08 & 3.51 \\
$E_\vK$ ($\mu$eV) & 117 & 68.5  & 79.4 &  28.9  \\

$T_K$ (K) & 1.36 &  0.8 & 0.92 & 0.34 \\[1ex]
\hline

\end{tabular}

\end{table}

\textit{\textbf{Main results}} --

\noindent $\bullet$  For \textit{\textbf{zero}} exciton center-of-mass wave vector, the $\hat{\Delta}$ difference is diagonal in the eightfold ground-exciton subspace, all states suffering the \textit{same} energy shift. So, the hole mass difference does not break the  exciton degeneracy. The effective average hole mass associated with the ground exciton binding, as obtained from Eq.~(\ref{WE_145_0}), reads 
\be\label{mh**}
m_h^{\ast\ast}\simeq
m_h^\ast \left[1+\beta 
\frac{ m_e}{m_e+m_h^\ast}
\left(\frac{m_{_H}-m_{_L}}{m_{_H}+m_{_L}}\right)^2\right]
\ee
 with $\beta \simeq0.90$. It deviates from $m^\ast_h$ at second order only in hole mass difference, with a larger deviation for large electron mass.  Table \ref{table1} for zinc-blende-like semiconductors, shows that the hole mass $m_h^{\ast\ast}$ is heavier than $m_h^\ast$ by $15\%$ to $20\%$.

\noindent $\bullet$  For \textit{\textbf{nonzero}} exciton wave vector $\vK$, the $\hat{\Delta}$ difference splits the eightfold ground exciton into two branches in a  tricky way: the resulting heavy and light excitons are not made of heavy or light holes as na\"{\i}vely thought, but of their spin-orbit-like combinations, $\mathcal{J}_\vK=\pm3/2$ and $\mathcal{J}_\vK=\pm1/2$, with $\mathcal{J}_\vK$  quantized along $\vK$. The associated center-of-mass energies, given in Eq.~(\ref{WE_148_0}), read  (Table \ref{table1})
\be\label{HL_5}
E_{\vK}^{(\pm)}=\frac{\hbar^2\vK^2}{2M_{_X}^{(\pm)}}\simeq
\frac{\hbar^2\vK^2}{2(m_e+m^\ast_h)}\left[1\pm \frac{m_{_H}-m_{_L}}{m_{_H}+m_{_L}}\,\,\frac{m^\ast_h}{m_e{+}m^\ast_h}\right]
\ee 
 When the hole mass ratio $m_{_L}/m_{_H}$ decreases from 1 to 0, the center-of-mass ratio $(m_e+m^\ast_h)/M_{_X}^{(\pm)}$ varies from 1 to $1\pm m^\ast_h/(m_e+m^\ast_h)=1\pm 2m_{_L}/(m_e+2m_{_L})$: the smaller the  electron mass compared to light hole, the larger the change in center of mass, a result hard to anticipate.

\textit{\textbf{Relevant Hamiltonians}} --
The Hamiltonian $\hat{H}$ for one electron and one hole reduces to the kinetic energies of the spatially nondegenerate electron and the threefold hole, plus their Coulomb interaction, $\hat{H}=\hat{K}^{(e)}+\hat{K}^{(h)}+\hat{V}^{(eh)}$. 

\noindent $\bullet$ The kinetic part for $m_e$ conduction electron reads
\be\label{HL_6}
\hat{K}^{(e)}=\sum_{\vk_e}\sum_{s=\pm1/2}\frac{\hbar^2\vk_e^2}{2m_e}\hat{a}^\dag_{\vk_e,s}\hat{a}_{\vk_e,s}
\ee
where $\hat{a}^\dag_{\vk_e,s}$ creates a $\vk_e$ electron with spin $s$ quantized along an arbitrary axis. 
 
  In the spherical approximation that neglects the warping, the upper fourfold valence level consists of holes having different masses $(m_{_H}, m_{_L})$, labeled by $\mathcal{J}_{\vk_h}=(\pm3/2,\pm1/2)$ index quantized along the hole wave vector $\vk_h$. The hole kinetic part reads
 \be
 \label{HL_7}
\hat{K}^{(h)}=\sum_{\vk_h}   \sum_{\mathcal{J}=(\pm3/2,\pm1/2)}\frac{\hbar^2\vk_h^2}{2m_{_{\mathcal{J}_{\vk_h}}}} \hat{b}^\dag_{\vk_h,\mathcal{J}_{\vk_h}}\hat{b}_{\vk_h,\mathcal{J}_{\vk_h}} 
\ee  
 where $\hat{b}^\dag_{\vk_h,\mathcal{J}_{\vk_h}}$ creates a heavy or light hole with mass $(m_{\pm3/2},m_{\pm1/2})=(m_{_H},m_{_L})$. 

 \noindent $\bullet$ The electron-hole Coulomb interaction is diagonal \cite{Monicbook} when written in terms of holes in Bloch states labeled by $\mu=(x,y,z)$, or in spin-orbit states labeled by $\mathcal{J}_\textbf{z}$ quantized along the crystal axis $\textbf{z}$\bea
\label{HL_8}
\hat{V}^{(eh)}&=& - \sum_{s=\pm1/2}  \,\,\, \sum_{\mathcal{J}=(\pm3/2,\pm1/2)}
 \\ 
&& 
\sum_{\vq\not=0} \frac{4\pi e^2}{\epsilon_{sc} L^3 q^2} \sum_{\vk_e\vk_h}
\hat{a}^\dag_{\vk_e+\vq,s_\textbf{z}}\hat{b}^\dag_{\vk_h-\vq,\mathcal{J}_\textbf{z}}\hat{b}_{\vk_h,\mathcal{J}_\textbf{z}}\hat{a}_{\vk_e,s_\textbf{z}}
\nn
\eea
for a sample volume $L^3$ and a dielectric constant $\epsilon_{sc}$. 

By contrast, this interaction does not stay diagonal, with complicated Coulomb scatterings\cite{Shiauprb21}, when written in terms of heavy/light holes with $\mathcal{J}_{\vk_h}$ index  quantized along the hole wave vector $\vk_h$. Indeed, the  creation operators for the two sets of holes are related by the following basis change 
\be\label{HL_9}
\hat{b}^\dag_{\vk_h,\mathcal{J}_{\vk_h}}=\sum_{\mathcal{J}'_\textbf{z}=(\pm3/2,\pm1/2)} \hat{b}^\dag_{\vk_h,\mathcal{J}'_\textbf{z}}\,\,\,\,{}_\textbf{z}\lan \mathcal{J}'| \mathcal{J}\ran_{\vk_h}
\ee 
where the ${}_\textbf{z}\lan \mathcal{J}'| \mathcal{J}\ran_{\vk_h}$ overlap depends\cite{Shiauprb21} on the $(\theta_{\vk_h},\varphi_{\vk_h})$ angles of the $\vk_h$ vector in the $(x,y,z)$ crystal axes, $\mathcal{J}_{\vk_h}$ reducing to $\mathcal{J}_\textbf{z}$ for $\vk_h$ along $\textbf{z}$.

\textit{\textbf{Exciton basis}} --
To derive the effects of the hole mass difference on exciton, we  introduce as exciton basis, the eigenstates of 
 \bea
 \label{12}
 \hat{H}_0=\hat{K}^{(e)}+\hat{K}^{(h)}_0+\hat{V}^{(eh)}
 \eea
where $\hat{K}^{(h)}_0$ reads as $\hat{K}^{(h)}$ with $(m_{_H},m_{_L})$ replaced by $m_h$,  which is  determined by minimizing the resulting exciton energy in the presence of different hole masses.

\noindent $\bullet$ To calculate the $ \hat{H}_0$ eigenstates, we first note that 
\be
\label{HL_14}
\sum_{\mathcal{J }=(\pm3/2,\pm1/2)}|\mathcal{J}\ran_{\vk_h} {}_{\vk_h}\lan \mathcal{J}|=\sum_{\mathcal{J}=(\pm3/2,\pm1/2)}|\mathcal{J}\ran_\textbf{z} {}_\textbf{z}\lan \mathcal{J}|
\ee
since they both form a complete set. This allows us to replace
$\mathcal{J}_{\vk_h}$ by $\mathcal{J}_\textbf{z}$ in $\hat{K}^{(h)}_0$, since the hole masses in $\hat{K}^{(h)}_0$ are the same. The $\hat{H}_0$ Hamiltonian then splits as
\be\label{HL_15}
\hat{H}_0=\sum_{s_\textbf{z}=\pm1/2}\,\,\,\sum_{\mathcal{J}_\textbf{z}=(\pm3/2,\pm1/2)}\hat{h}_{s_\textbf{z},\mathcal{J}_\textbf{z}}
\ee
where $\hat{h}_{s_\textbf{z},\mathcal{J}_\textbf{z}}$ is a hydrogen-like Hamiltonian. 

\noindent $\bullet$ The resulting eigenenergies split  into a center-of-mass part and a relative-motion part 
\be\label{HL_17}
E_{\vK,\nu;s_\textbf{z},\mathcal{J}_\textbf{z}}=\frac{\hbar^2\vK^2}{2(m_e+m_h)}+\va_\nu
\ee
 The creation operators for the corresponding $(2\times4)$-fold excitons $|\vK,\nu;s_\textbf{z},\mathcal{J}_\textbf{z}\ran$ read as
\be\label{HL_19}
\hat{B}^\dag_{\vK,\nu;s_\textbf{z},\mathcal{J}_\textbf{z}}=\sum_\vk \hat{a}^\dag_{-\vk+\gamma_e\vK,s_\textbf{z}}\hat{b}^\dag_{\vk+\gamma_h\vK,\mathcal{J}_\textbf{z}}\lan \vk|\nu\ran
\ee 
 where $\lan \vk|\nu\ran$ is the hydrogen-like  wave function, with energy $\va_\nu$ that scales in Rydberg unit\cite{Landau}
\be\label{HL_18}
R_{_X}=\frac{\mu_{_X} e^4}{2\hbar^2 \epsilon_{sc}^2}=\frac{\hbar^2}{2\mu_{_X} a_{_X}^2}
\ee
for $\mu_{_X}^{-1}=m_e^{-1}+m_h^{-1}$.

These $\hat{H}_0$ eigenstates are used as a basis to approach excitons with different hole masses.

\textit{\textbf{Effects of heavy and light holes}} --
The hole mass difference is concentrated into the one-body operator 
\bea
 \label{12'}
 \hat{\Delta}=\hat{H}-\hat{H}_0=\hat{K}^{(h)}-\hat{K}^{(h)}_0
 \eea

\noindent $\bullet$ To better catch its effect in the $\hat{H}_0$ exciton subspace, we split $\hat{\Delta}$ as $\hat{\Delta}^{(h)}+\hat{\Delta}^{(HL)}$. The part $\hat{\Delta}^{(h)}=\hat{K}_0^{(h^*)}-\hat{K}_0^{(h)}$ for $\hat{K}_0^{(h^*)}$ reading as $\hat{K}_0^{(h)}$ with $m_h$ replaced by $m^*_h$, concentrates on using $m_h$ instead of the  commonly accepted $m^*_h$ value. The part $\hat{\Delta}^{(HL)}=\hat{K}^{(h)}-\hat{K}_0^{(h^*)}$, which cancels for $m_{_H}=m_{_L}$, concentrates on the mass difference.

\noindent $\bullet$ The $\hat{\Delta}$ matrix elements in the $\nu=\nu_0$ ground exciton subspace
\be\label{HL_24}
\lan \vK',\nu_0;s'_\textbf{z},\mathcal{J}'_\textbf{z}|\hat{\Delta}|\vK,\nu_0;s_\textbf{z},\mathcal{J}_\textbf{z}\ran=\delta_{\vK',\vK}\delta_{s'_\textbf{z},s_\textbf{z}}\Delta_{\vK;\mathcal{J}'_\textbf{z},\mathcal{J}_\textbf{z}}
\ee
 differ from zero for $s'_\textbf{z}=s_\textbf{z}$ and $\vK'=\vK$ because $\hat{\Delta}$, that comes from difference in hole kinetic energies, does not act on spin nor on wave vector. This feature holds true for higher-order terms in $\hat{\Delta}$.

\textit{\textbf{Change in binding energy}} --
We first focus on binding energy, \textit{i.e.,} $\textbf{K}=\bf0$ exciton.

 \noindent $\bullet$ We find that the first-order contribution in $\hat{\Delta}^{(HL)}$ cancels (Eq.~(S39) of \cite{SM}),
  \be
\label{34}
\Delta^{(HL)}_{\textbf{0};\mathcal{J}'_\textbf{z},\mathcal{J}_\textbf{z}}=0
\ee
while the $\hat{\Delta}^{(h)}$ part leads to (Eq.~(S41) of \cite{SM})
\be
\Delta^{(h)}_{\textbf{0};\mathcal{J}'_\textbf{z},\mathcal{J}_\textbf{z}}= \delta_{\mathcal{J}'_\textbf{z},\mathcal{J}_\textbf{z}} \frac{\hbar^2}{2 a_{_X}^2}\left(\frac{1}{m_h^*}-\frac{1}{m_h}\right)\label{HL_27}
\ee
Being diagonal in $\mathcal{J}_\textbf{z}$, the $\Delta^{(h)}$ difference thus produces the same energy shift 
\be
R_{_X} \mu_{_X}\left(\frac{1}{\mu^*_{_X}}-\frac{1}{\mu_{_X}}\right)
\ee
to all eightfold ground-exciton states, that is, no splitting. The resulting exciton energy then reads
  \be
  \label{23'}
R_{_X}\left[-1+ \mu_{_X}\left(\frac{1}{\mu^*_{_X}}-\frac{1}{\mu_{_X}}\right)\right]=R_{_X}^*\frac{\mu_{_X}}{\mu^*_{_X}}\left[-2+\frac{\mu_{_X}}{\mu^*_{_X}}\right]
\ee
 where $R_{_X}^*$ is given by Eq.~(\ref{HL_18}) for $m_h=m_h^*$. Its minimum value, obtained for $\mu_{_X}=\mu^*_{_X}$, gives a maximum ground-exciton binding equal to $R_{_X}^*$. So, up to first order in   $\hat{\Delta}$, the hole mass difference does not break the ground-exciton  degeneracy, the appropriate hole mass value for binding being the one given in Eq.~(\ref{HL_2}).

\noindent $\bullet$ Similar but somewhat heavier calculations\cite{SM} performed for the product $\hat{\Delta}^{(h)}\,\hat{\Delta}^{(h)}$, show  that the $(m_h-m^*_h)$ difference leads to (Eq.~(S65) of \cite{SM})
\be
\label{WE_140}
\Delta^{(h,h)}_{{\bf0},\mathcal{J}'_\textbf{z},\mathcal{J}_\textbf{z}}=\delta_{\mathcal{J}'_\textbf{z},\mathcal{J}_\textbf{z}}
\left[ \frac{\hbar^2}{2a_{_X}^2}
\left(\frac{1}{m_h^*}-\frac{1}{m_h}\right) \right]^2
\frac{ \beta_{{_1}}}{ R_{_X}   }
\ee
 with $\beta_{_1}\simeq-0.50$. In contrast to first order, the $(m_{_H}-m_{_L})$ difference now brings a nonzero contribution through $\hat{\Delta}^{(HL)}\,\hat{\Delta}^{(HL)}$, equal to (Eq.~(S56) of \cite{SM})
\be
\label{WE_137}
\Delta^{(HL,HL)}_{{\bf0},\mathcal{J}'_\textbf{z},\mathcal{J}_\textbf{z}}=\delta_{\mathcal{J}'_\textbf{z},\mathcal{J}_\textbf{z}}\left[ \frac{\hbar^2}{2a_{_X}^2}
\left(\frac{1}{m_{_H}}-\frac{1}{m_{_L}}\right) \right]^2
\frac{ \beta_{{_2}}}{ R_{_X}   }\ee
with $\beta_{_2}\simeq -0.22$, while the mixed term in $\hat{\Delta}^{(HL)\,}\hat{\Delta}^{(h)}$ gives no contribution.

\noindent $\bullet$ Since the above contributions also are diagonal in $\mathcal{J}_\textbf{z}$, the exciton degeneracy remains unbroken, but the energy shift they produce now depends on the hole mass difference. When combined with Eq.~(\ref{23'}), the exciton energy now reads as
\bea
R_{_X} \left[-1+\mu_{_X} \left(\frac{1}{\mu_{_X}^*}{-}\frac{1}{\mu_{_X}}\right) +\mu_{_X}^2\left(\frac{1}{\mu_{_X}^*}{-}\frac{1}{\mu_{_X}}\right)^2 \beta_1\right.\nn\\
\left.+\mu_{_X}^2 \left(\frac{1}{m_{_H}}{-}\frac{1}{m_{_L}}\right)^2 \beta_2 \right]\hspace{1cm}\label{WE_143}
\eea 
Its minimum value, obtained for 
\be
\mu_{_X}=\mu_{_X}^*
\left(1- \frac{3\beta_2}{2+2\beta_1}\mu_{_X}^{\ast\,2}\left(\frac{1}{m_{_H}}-\frac{1}{m_{_L}}\right)^2 \right)
\label{WE_145}
\ee
 gives the ground exciton energy as (Eq.~(S71) of \cite{SM})
\be\label{WE_145_0}
 R_{_X}^*\left[-1+\beta_2 \mu_{_X}^{\ast\,2}\left(\frac{1}{m_{_H}}-\frac{1}{m_{_L}}\right)^2\right]
\equiv - \frac{\mu_{_X}^{\ast\ast} e^4}{2\hbar^2 \epsilon_{sc}^2}
\ee
that corresponds, for $\mu_{_X}^{\ast\ast \,-1}=m_e^ {-1}+m_h^{\ast\ast\,-1}$, to the hole effective mass $m_h^{\ast\ast}$ given in Eq.~(\ref{mh**}).

\textit{\textbf{Change in center-of-mass  energy}} --
We now look for the effect of the $\hat{\Delta}$ difference on the eightfold ground exciton subspace when its center-of-mass wave vector $\vK$ differs from zero. This will tell how the hole mass difference affects the exciton  motion. By taking $m_h=m_h^*$, which produces the correct binding  at first order in $\hat{\Delta}$, this difference reduces to $\hat{\Delta}^{(HL)}$.  Its contribution to the exciton energy appears as (Eq.~(S73) of \cite{SM})
\be
\label{HL_41}
\Delta^{(HL)}_{\vK;\mathcal{J}'_\textbf{z},\mathcal{J}_\textbf{z}}=\sum_{\vk} |\lan\vk-\gamma^\ast_h\vK |\nu_0\ran|^2 \frac{\hbar^2\vk^2}{2m_{_{HL}}}\, {}_\textbf{z}\lan \mathcal{J}'|\hat{D}_{\vk}|\mathcal{J}\ran_\textbf{z}
\ee
 with $\gamma^\ast_h=  m_h^\ast /(m_e+m_h^\ast)$ and $1/ m_{_{HL}}=(1/m_{_H}-1/m_{_L})/2$,  for  $\hat{D}_\vk$ given by
\bea
\label{30}
\hat{D}_\vk=\Big(\sum_{\mathcal{J}=\pm3/2}-\sum_{\mathcal{J}=\pm1/2} \Big) |\mathcal{J}\ran_\vk\,{}_\vk\lan  \mathcal{J}|
\eea

As $\Delta^{(HL)}_{\vK;\mathcal{J}'_\textbf{z},\mathcal{J}_\textbf{z}}$ is equal to zero for $\vK=\v0$, its lowest-order term in $\vK$, obtained from the $|\lan\vk-\gamma^\ast_h\vK |\nu_0\ran|^2$ expansion, leads after summing over $\vk$, to
\be
\label{WE_148}
\Delta^{(HL)}_{\vK;\mathcal{J}'_\textbf{z},\mathcal{J}_\textbf{z}}\simeq \frac{\hbar^2\gamma^{\ast\,2}_h \vK^2}{2 m_{_{HL}}}\, {}_\textbf{z}\lan \mathcal{J}'|\hat{D}_{\vK}|\mathcal{J}\ran_\textbf{z}
\ee
To go further, we note that changing the hole quantization axis from $\textbf{z}$ to $\vK$, renders diagonal the above matrix elements. This leads to (Eq.~(S80) of \cite{SM})
\be\label{WE_148_0}
\Delta^{(HL)}_{\vK;\mathcal{J}'_\vK,\mathcal{J}_\vK}=\delta_{\mathcal{J}'_\vK,\mathcal{J}_\vK}\frac{\hbar^2\gamma_h^{\ast \,2}\vK^2}{2m_{_{HL}}}\Big(\delta_{\mathcal{J}_\vK,\pm\frac3 2}{-}\delta_{\mathcal{J}_\vK,\pm\frac 1 2}\Big)
\ee

  For different hole masses, the exciton center-of-mass energy thus splits into a heavy branch, $\mathcal{J}_\vK=\pm3/2$, and a light branch, $\mathcal{J}_\vK=\pm1/2$, as said in Eq.~(\ref{HL_5}), the energy difference being more significant for light electron, $m_e\ll m^\ast_h$. Note that these heavy and light exciton masses differs from the na\"{\i}ve $(m_e+m_{_H})$ and $(m_e+m_{_L})$ values the excitons would have if they were simply made of heavy or light hole.
  

\textit{\textbf{Experimental consequences}} --

\noindent $\bullet$ The most direct way to evidence this exciton splitting would be to see two distinct absorption lines associated with the two exciton branches.  This appears difficult in the case of GaAs because the exciton linewidth in usual samples  ($\simeq320\,\mu$eV in \cite{Gopal2000}) is  large compared with the $E_{\vK}^{(\pm)}$ difference $(\simeq117\,\mu$eV in Table \ref{table1}), the photocreated exciton having a wave vector equal to the photon wave vector; so, $E_{gap}\simeq\hbar (c/n_r) |\vK|$ where $n_r$ is the semiconductor refractive index. As a result, the heavy and light exciton lines would appear as a single broad line.

\noindent $\bullet$
Another idea is to note that due to their energy differences, the thermal populations of heavy and light excitons evolve differently 
\be
N_{\vK,T}^{(\pm)}=N_{\vK,\infty}\exp(-E_{\vK}(1\pm r)/k_{_B}T)\propto \exp(\pm rT_K/T)
\ee 
for $T_K=E_{\vK}/k_{_B}\simeq 1.4\,$K. Since heavy excitons have a lower energy, they contribute to the low-energy side of the absorption line, with a number that is not only larger than the light exciton number but that also increases faster when $T$ decreases; so, the shape of the absorption line should show an enhancement on the low-energy side.

Another way to quantify the difference in the heavy and light exciton populations, is through the total weight of the photoluminescence line as a function of temperature: being an overlap of two lines with amplitudes that vary differently, the $\exp(2rT_K/T)$ difference must show up in a Log plot of the temperature evolution in the subKelvin temperature range.  

 \noindent $\bullet$
 Such temperature dependences can be directly compared to the predicted exciton splitting driven by heavy-light hole Coulomb couplings, provided that there is no extrinsic effect associated with residual crystal strain\cite{1,3,5}. Regarding this point, it has been shown\cite{7} that epitaxial GaAs samples on a Si substrate are essentially strain-free beyond a critical thickness $\simeq2.3\, \mu$m.
 
Another major problem is that, for such low temperatures, bright excitons are going to mostly condense into dark states\cite{BEC,dubin2017}, making their observation quite stringent.

\textit{\textbf{Conclusion}} -- We here study the consequences of valence holes having two different masses, ($m_{_{H}},m_{_{L}}$), on the exciton energy, due to their nondiagonal intraband Coulomb coupling\cite{earlywork}.  

We show that for zero center-of-mass wave vector, the ground-state exciton stays $(2\times4)$-fold degenerate, its binding energy reading in terms of a \textit{single} hole mass that is close to the commonly-used average value $m_h^\ast$ given in Eq.~(\ref{HL_2}). We expect this conclusion, mathematically proved up to second order in hole mass difference, to stay valid up to any order for a fundamental reason: in the absence of exciton wave vector, there is no symmetry-breaking axis to possibly split the exciton degeneracy. This strong argument leads us to anticipate that, when included, the valence band warping will not  split the exciton energy\cite{warping}.

By contrast, a nonzero center-of-mass wave vector $\textbf{K}$ splits the exciton degeneracy into a heavy branch and a light branch, with $(\pm3/2)$ and $(\pm1/2)$ indices quantized along the symmetry-breaking axis $\textbf{K}$. The associated center-of-mass masses depend on the ($m_{_{H}}-m_{_{L}}$) difference and the electron mass $m_{e}$ as given in Eq.~(\ref{HL_5}).

This splitting raises fundamental questions that deserve further investigation on using a unique hole mass for many-body effects in which excitons with finite wave vector play a key role, like the exciton Bose-Einstein condensation\cite{BEC,BEC2,BEC3}, the two-component condensates\cite{Kuklov,Cabrera}, and the BEC-BCS crossover\cite{BEC-BCS,Li}.


\begin{thebibliography}{99}
\bibitem{earlywork} Early works\cite{Baldereschi_prl,Baldereschi_prb} that mainly focus on exciton binding corrections do not address these consequences because they stay with the effective-mass Hamiltonian, instead of the spherical Hamiltonian.
\bibitem{warping}The effect of the hole warping on the exciton binding and center-of-mass motion will be considered elsewhere.
\end{thebibliography}
\end{document}